\documentclass[aps,prb,preprint,showpacs,floatfix]{revtex4}

\usepackage{amsmath}
\usepackage{subfigure}

\newif\ifpdf
\ifx\pdfoutput\undefined
\pdffalse 
\else
\pdfoutput=1 
\pdftrue
\fi

\ifpdf
\usepackage[pdftex]{graphicx}
\DeclareGraphicsExtensions{.pdf,.jpg,.tif}
\else
\usepackage{graphicx}
\DeclareGraphicsExtensions{.eps,.jpg,.ps}
\fi

\begin{document}

\title{Classical and quantum kinetics of the Zakharov system}

\author{Mattias Marklund}

\affiliation{Department of Physics, Ume{\aa} University, SE--901 87 Ume{\aa},
  Sweden}
  
\date{\today}

\begin{abstract}
  A kinetic theory for quantum Langmuir waves interacting nonlinearly 
  with quantum ion-acoustic waves is derived. The formulation allows
  for a statistical analysis of the quantum correction to the Zakharov 
  system. The influence of a background random phase on the
  modulational instability is given. In the coherent case, the effect of 
  the quantum correction is to reduce the growth rate. Moreover, in the classical limit, a
  bifurcation develops in the dispersion curves due to the presence of 
  partial coherence. However, the combined effect of partial coherence and 
  a quantum correction may give rise to an increased modulational instability 
  growth rate, as
  compared to the classical case. The results may be of significance 
  in dense astrophysical plasmas and laboratory laser--plasma systems.
\end{abstract}

\maketitle

\section{Introduction}

The effects from the quantum domain have intriguing consequences for the
way we view the world and the way we interpret physical models, and 
these effects can be seen in single particle phenomena. Plasma
physics, on the other hand, deals with the collective interaction of charged
particles, and processes related to such. However, there are physical 
parameter domains where quantum mechanics and plasma physics
need to be taken into account simultaneously, e.g. dense and/or hot
astrophysical plasmas. Indeed, the creation of pair plasmas surrounding
neutron stars can be viewed as a collective quantum plasma effect. Thus,
there is interest and application for a model taking both collective 
charged particle effects and quantum phenomena into account (see, e.g.
Refs.\ \onlinecite{Haas-etal} and \onlinecite{Hall-etal}).

One of the most prominent models in plasma physics is described
by the Zakharov equations \cite{Zakharov}, in which high frequency Langmuir waves  
are coupled nonlinearly to low frequency ion-acoustic waves.   
The statistical properties of this system has been analyzed in Ref.\
\onlinecite{Fedele-etal}, where a Landau-like damping was found. 
Recently, a generalization of the Zakharov system was derived,
taking quantum effects into account \cite{Garcia-etal}. The effect
of the quantum correction was to introduce higher order dispersion into
the system of equations, thus altering the behavior of wave evolution.
It was argued in Ref. \onlinecite{Garcia-etal} that these contributions
could be important in astrophysical plasmas, as the plasma densities
may become significant. 

In this paper, we will introduce a kinetic description of the 
quantum Zakharov equation, by applying the Wigner transform to
the Langmuir propagation equation. The resulting system of equations
may be useful for understanding the properties of partially 
coherent Langmuir waves interacting with quantum ion-acoustic waves.
We derive the general dispersion relation, and analyze the stability of the system. 
A comparison between monoenergetic Langmuir waves and 
random-phase Langmuir waves is given, and it is found that the 
interplay between quantum corrections and spectral broadening
may alter the instability properties in novel ways. In particular,
it is found that the growth rate for a short wavelength partially coherent quantum Langmuir
wave is larger than the corresponding growth rate for the classical Langmuir wave.
Thus, the interplay between incoherence and quantum effects gives rise
to modified modulational instability growth rates, a result that may be 
relevant to astrophysical and intense laboratory laser-plasmas, for which the 
quantum parameter may take on significant values.

\section{Basic equations}

The dynamics of the nonlinearly coupled quantum Langmuir and ion-acoustic waves 
is given by the Zakharov-like equations \cite{Garcia-etal}
\begin{subequations}
\begin{equation}\label{eq:langmuir}
  i\partial_tE(t,x) + \partial^2_xE(t,x) - H^2\partial^4_xE(t,x) = n(t,x)E(t,x) ,
\end{equation}
and
\begin{equation}\label{eq:ion}
  (\partial^2_t - \partial^2_x)n(t,x) + H^2\partial^4_xn(t,x) = \partial^2_x|E(t,x)|^2 ,
\end{equation}
\label{eq:eqs}
\end{subequations}
where $H \equiv \hbar\omega_{pi}/k_bT_e$ is the quantum parameter
due to a quantum pressure emanating from the underlying hydrodynamic 
model \cite{Garcia-etal}. 
Here $\hbar$ is Planck's constant divided by $2\pi$, 
$\omega_{pi} = (n_0e^2/m_i\epsilon_0)^{1/2}$ is the ion plasma
frequency, 
$k_B$ is Boltzmann's constant, $T_e$ is the electron temperature,
$n_0$ is the constant background density, and $m_i$ is the ion rest mass.
The electric field $E$ has been normalized according to
$E \rightarrow (\epsilon_0m_i/16m_en_0k_BT_e)^{1/2}E$, while
the density $n$ is normalized by $n \rightarrow (m_i/4m_en_0)n$, 
where $m_e$ is the electron mass. The coordinates have been rescaled
using $t \rightarrow (2m_e/m_i)\omega_{pe}t$ and $x \rightarrow 
2(m_e/m_i)^{1/2}x/\lambda_e$, where $\omega_{pe} = 
(n_0e^2/m_e\epsilon_0)^{1/2}$ is the electron plasma frequency and
$\lambda_e$ is the electron Debye length. 
As $H \rightarrow 0$, we regain the classical Zakharov equations 
from (\ref{eq:eqs}). However, in some astrophysical 
plasmas, the quantum parameter $H$ may approach unity,
since in such environments, high densities are not uncommon
(see, e.g. Ref.\ \onlinecite{Opher-etal}). We see that the effect of the quantum
parameter is to introduce higher order dispersion.

\section{Quantum kinetics}

The Fourier transform of the two-point correlation function, as
given for the electric field by
\begin{equation}\label{eq:wigner}
  \rho(t,x,p) = \frac{1}{2\pi}\int\,d\xi\,e^{ip\xi}
    \langle E^*(t, x + \xi/2)E(t, x - \xi/2)\rangle
\end{equation}
was introduced by Wigner \cite{Wigner} in quantum statistical
mechanics. Here the angular brackets denotes the ensemble average, 
and the asterisk denotes the complex conjugation operation.
The Wigner function $\rho$ is a generalized distribution
function, which satisfies
\begin{equation}\label{eq:norm}
  \langle|E(t,x)|^2\rangle = \int\,dp\,\rho(t,x,p) .
\end{equation} 
Applying the transformation (\ref{eq:wigner}) to Eq. (\ref{eq:langmuir})
gives the kinetic equation \cite{Helczynski-etal}
\begin{equation}\label{eq:kinetic}
  \partial_t\rho(t,x,p) 
  + \left( 2p\partial_x + 4H^2p^3\partial_x - H^2p\partial^3_x \right)\rho(t,x,p)
  - 2n(t,x)\sin\left( \tfrac{1}{2}\stackrel{\leftarrow}{\partial_x}
    \stackrel{\rightarrow}{\partial_p} \right)\rho(t,x,p) = 0 ,
\end{equation}
which is coupled to the ion-acoustic equation (\ref{eq:ion}) via
Eq. (\ref{eq:norm}). Here the $\sin$-operator is defined
by its Taylor expansion, and arrows denote direction 
of operation. Keeping the lowest order derivative in this Taylor
expansion, corresponding to the long wavelength limit,  
gives a modified Vlasov equation 
\begin{equation}
  \partial_t\rho(t,x,p) 
  + pD_x\rho(t,x,p)
  - (\partial_xn(t,x))({\partial_p}\rho(t,x,p)) = 0 ,
\end{equation}
for the quantum Langmuir wave, driven by the ion-acoustic
ponderomotive force. Here
$D_x \equiv (2 + 4H^2p^2 - H^2\partial^2_x)\partial_x$. Thus,
in the classical limit $H \rightarrow 0$, $D_x \rightarrow 2\partial_x$, and 
we obtain a Vlasov-like equation for the
long wavelength Langmuir waves. 

\section{The modulational instability} 

In order to analyze Eqs. (\ref{eq:ion}), (\ref{eq:norm}), and (\ref{eq:kinetic}),
we perform a perturbative expansion. Letting $\rho(t,x,p) = 
\rho_0(p) + \rho_1\exp(ikx - i\omega t)$, where $|\rho_1| \ll \rho_0$,
and $n(t,x) = n_0 + n_1\exp(ikx - i\omega t)$, we linearize with respect
to the perturbation variables. We then obtain the dispersion relation
\begin{eqnarray}
  -\omega^2 + (1 + H^2k^2)k^2 =
  k^2\int\,dp\,\frac{\rho_0(p + k/2) - \rho_0(p - k/2)}{\omega - 
    kp(2 + 4H^2p^2 + H^2k^2)} .
\label{eq:dispersion}
\end{eqnarray}
The dispersion relation (\ref{eq:dispersion}) generalizes the results in 
Refs.\ \onlinecite{Fedele-etal} and 
\onlinecite{Garcia-etal}, and is valid for partially coherent 
quantum Langmuir waves interacting
nonlinearly with quantum ion-acoustic waves.

\subsection{Monoenergetic Langmuir waves}

In the case of a monoenergetic Langmuir wave, we have 
$\rho_0(p) = I_0\delta(p)$,, where $I_0 = |E_0|^2$ is the
background intensity. Then the dispersion relation (\ref{eq:dispersion})
becomes
\begin{equation}\label{eq:monodisp}
  \left[ \omega^2 - (1 + H^2k^2)k^2 \right]\left[ \omega^2  - (1 + H^2k^2)^2k^4 \right]
  = 2I_0(1 + H^2k^2)k^4 , 
\end{equation}
such that
\begin{equation}
  \omega^2 = \tfrac{1}{2}\left[ \bar{H}^2k^2 + \bar{H}^4k^4 \pm 
    \bar{H}k^2\sqrt{\bar{H}^2 + 8I_0 - 2\bar{H}^4k^2 + \bar{H}^6k^4}\,\right] ,
\end{equation}
where $\bar{H} \equiv 1 + H^2k^2$. Letting $\omega = i\gamma$, the
instability growth rate is given by \cite{Garcia-etal}
\begin{equation}\label{eq:growthrate}
  \gamma = \tfrac{1}{\sqrt{2}}\left[ \bar{H}k^2\sqrt{\bar{H}^2 + 8I_0 - 2\bar{H}^4k^2 + \bar{H}^6k^4}\,
    - \bar{H}^2k^2 - \bar{H}^4k^4 \right]^{1/2} .
\end{equation} 
Starting from $H = 0$, successively higher values of $H$ tend to suppress
the instability, giving lower growth rates with a cut-off at a lower wavenumber,
see Fig.\ \ref{fig:1}. 

\subsection{Partial coherence}

The coherent monoenergetic background distribution 
gives important information on wave instabilities. 
However, in many applications the background field is not fully coherent, 
but rather displays partial decoherence due to, e.g.\ noise. The noise, either
classical or quantum, may stem from different sources, such as thermal 
effects, weak turbulence, or quantum fluctuations. Such sources of noise 
may lead to a background field $E_0$ with a random phase $\varphi(x)$ 
such that
\begin{equation}
  \langle e^{-i[\varphi(x + \xi/2) - \varphi(x - \xi/2)]}\rangle = 
  e^{-p_W|\xi|} ,
\end{equation} 
with the corresponding distribution function $\rho_0$ is given by the Lorentzian 
\cite{Loudon,Mendonca}
\begin{equation}
  \rho_0(p) = \frac{I_0}{\pi}\frac{p_W}{p^2 + p_W^2},
\end{equation}
where $p_W$ is the width of the distribution. The integrand of (\ref{eq:dispersion})
has three poles, one real and two complex, where the real pole is given by 
\begin{equation}
  p_0 = -\frac{1}{A^{1/3}(\omega,k)} + \frac{A^{1/3}(\omega,k)}{3b(k)} .
\end{equation}
Here $A(\omega,k) = 3[9ab^2 + \sqrt{3}(4b^3 + 27a^2b^4)^{1/2}]/2$, 
$a(\omega,k) = \omega/k(2 + H^2k^2)$, and $b(k) = 4H^2/(2 + H^2k^2)$. 
As the quantum parameter $H$ 
approaches zero, the complex poles approaches complex infinity. 
Thus, in the integration of Eq.\ (\ref{eq:dispersion})
we will neglect these poles, only taking the real pole $p_0$ into account, since the modes
corresponding to the complex poles are quickly damped. Thus, we have
\begin{eqnarray}
  && -\omega^2 + (1 + H^2k^2)k^2     =
  2I_0k^4 \Bigg\{ 
  \frac{g(k) - h(k)}%
    {\left[ \omega + 2ip_W h(k) k\right]^2 
    - k^4[g(k) - h(k)]^2} 
\nonumber \\ &&\qquad\qquad
   + \frac{ip_Wp_0}{k^2(2 + H^2k^2)[\left( 
   p_0 + k/2 \right)^2 + p_W^2][\left( 
   p_0 - k/2 \right)^2 + p_W^2]}
  \Bigg\}
\label{eq:newdisprel}
\end{eqnarray}
where we have defined the real and positive 
functions $g(k) = H^2(k^2 + 8p_W^2)$ and $h(k) = 1 +2H^2(k^2 + p_W^2)$. 
The dispersion relation (\ref{eq:newdisprel}) describes the effects of partial
coherence for the quantum Zakharov system (\ref{eq:eqs}). The damping
character due to the finite width of the background distribution can clearly be 
seen, as well as the Landau damping due to the real pole. We note that
as $p_W \rightarrow 0$, we regain the monoenergetic dispersion relation
(\ref{eq:monodisp}).


\subsection{The classical limit}

If $H \rightarrow 0$, we obtain the classical limit of the
dispersion relation (\ref{eq:newdisprel}), when we use a kinetic photon
description for the Langmuir waves. The effects of statistical broadening
on the Zakharov system was also analyzed in Ref.\ \onlinecite{Fedele-etal}.
We note that the 
two complex poles approaches infinity, and only 
the real pole remains with the value $p_0 = \omega/2k$, as it should.
The dispersion relation then reads \cite{Fedele-etal}
\begin{equation}\label{eq:classical}
 (\omega^2 - k^2)\left[\omega^4 + k^8 + 8p_W^2k^6 + 8p_W^2k^2\omega^2
   + 2k^4(8p_W^4 - \omega^2) \right] 
 = 2I_0k^4(\omega^2 - k^4 - 8ip_Wk\omega - 4p_W^2k^2) ,
\end{equation} 
and although the quantum effects have been neglected, the behavior 
of the function $\omega(k; p_W)$ is still rather complicated. The 
dispersion relation (\ref{eq:dispersion}) with $H = 0$ was analyzed analytically in the
long wavelength limit, i.e.\ $\omega/k \gg 1$, in Ref.\ \onlinecite{Fedele-etal},
and the growth rate was found for a Gaussian background spectrum. 
Here we will solve the equation (\ref{eq:classical}) for all wave lengths.   
Letting $\omega = \mathrm{Re}\,\omega + i\gamma$, we may
solve Eq.\ (\ref{eq:newdisprel}) numerically for the growth rate $\gamma$, 
using different values of the widths  
$p_W$. In the Figs.\ \ref{fig:classical} and \ref{fig:classical-large} 
we have plotted the solutions
for $H = 0$ and a number of values of $p_W$. The results show on
a more complicated dispersion structure than in the coherent case. 
The asymptotic behavior of the growth rate for short wavelengths has been
depicted in Fig.\ \ref{fig:wavelength}, using a number of different values on the
decoherence width $p_W$. For large $k$, the growth rate has a linear
dependence on the wavenumber, the slope being determined by the 
values of the width $p_W$.

\subsection{Quantum effects on the instability growth rate}

When $H$ is nonzero, the combined effects of quantum correction and
decoherence make themselves 
apparent in the dispersion relation (\ref{eq:newdisprel}) through 
new and novel wave modes. In Fig.\ (\ref{fig:quantum}) we display the 
growth rate $\gamma$
as a function of the wave number $k$. As compared to the classical case,
the combined effect of partial coherence and quantum 
effects, i.e.\ finite $p_W$ and $H$ respectively, is to make the modulational 
instability growth rate smaller for long wavelength perturbations. However,
the interesting effect is for short wavelengths, where the modes introduced 
due the finite spectral width is amplified by the quantum corrections. Thus,
we may expect much stronger growth rates for short wavelength perturbations,
making these dominant in quantum plasmas. 
In Fig.\ \ref{fig:quantum} the
dispersion curves for a value $H = 0.25$ of the quantum parameter has been
plotted. The strong growth rate for short wavelengths can clearly be seen.

\section{Conclusions}

The effects of partial coherence in quantum plasmas, such as in the
form of a random phase, is of interest in certain plasmas, such as 
astrophysical plasmas \cite{Opher-etal,Garcia-etal}, and the next generation
laser plasma systems \cite{Bulanov-etal,mou05}. Moreover, in such system, 
the density may even reach values of \cite{Opher-etal,Garcia-etal} 
$10^{23} - 10^{31}\,\mathrm{m}^{-3}$
for temperatures of the order $10^5 - 10^7\,\mathrm{K}$,
giving $H = \hbar\omega_{pi}/k_BT_e \sim 10^{-7} - 1$. 
Thus, the quantum 
parameter $H$ may attain appreciable values, 
such that the higher order dispersive terms 
in Eqs.\ (\ref{eq:eqs}) become important. Even in the cases of 
small a quantum correction, this effect combined with Langmuir wave 
decoherence will yield a strongly growing mode for short wavelengths, 
and could lead to significant changes in extreme astrophysical and
laboratory plasmas. Thus, the combination 
of incoherence and quantum effects may yield rich and interesting
dynamics of Langmuir wave propagation in such plasmas. However, a 
detailed analysis of possible applications is left for future research. 

Here we have analyzed the statistical properties of the quantum Zakharov
system, giving the dynamics of high frequency Langmuir waves 
in terms of a kinetic equation. This enabled the investigation into 
the effects of partial coherence of the quantum Langmuir wave, 
in particular the implications due to a random phase, and it was found that
such a system exhibits an interesting dispersion structure. In particular,
the combined effect of decoherence and quantum corrections gives rise
to new dispersion curves as well as increased modulational instability
growth rates, as compared to the case of a classical coherent and partial coherent 
Langmuir wave.

\acknowledgments
This research was supported by the Swedish Research Council 
through the contract No. 621-2004-3217. The author would like to
thank K.\ Markstr\"om for stimulating discussions and 
insights into algebraic computing. 


\newpage

\noindent Fig. 1: The growth rate $\gamma$ as given by Eq. (\ref{eq:growthrate})
    plotted as a function of the wavenumber $k$. Seen from the right, the values
    $H= 0,\, 0.5$ and $1$ have been used together with the intensity $I_0 = 0.5$.
    The effect of the quantum parameter is thus to suppress the instability. \\[3mm]
    
\begin{figure}
  \includegraphics[width=3in]{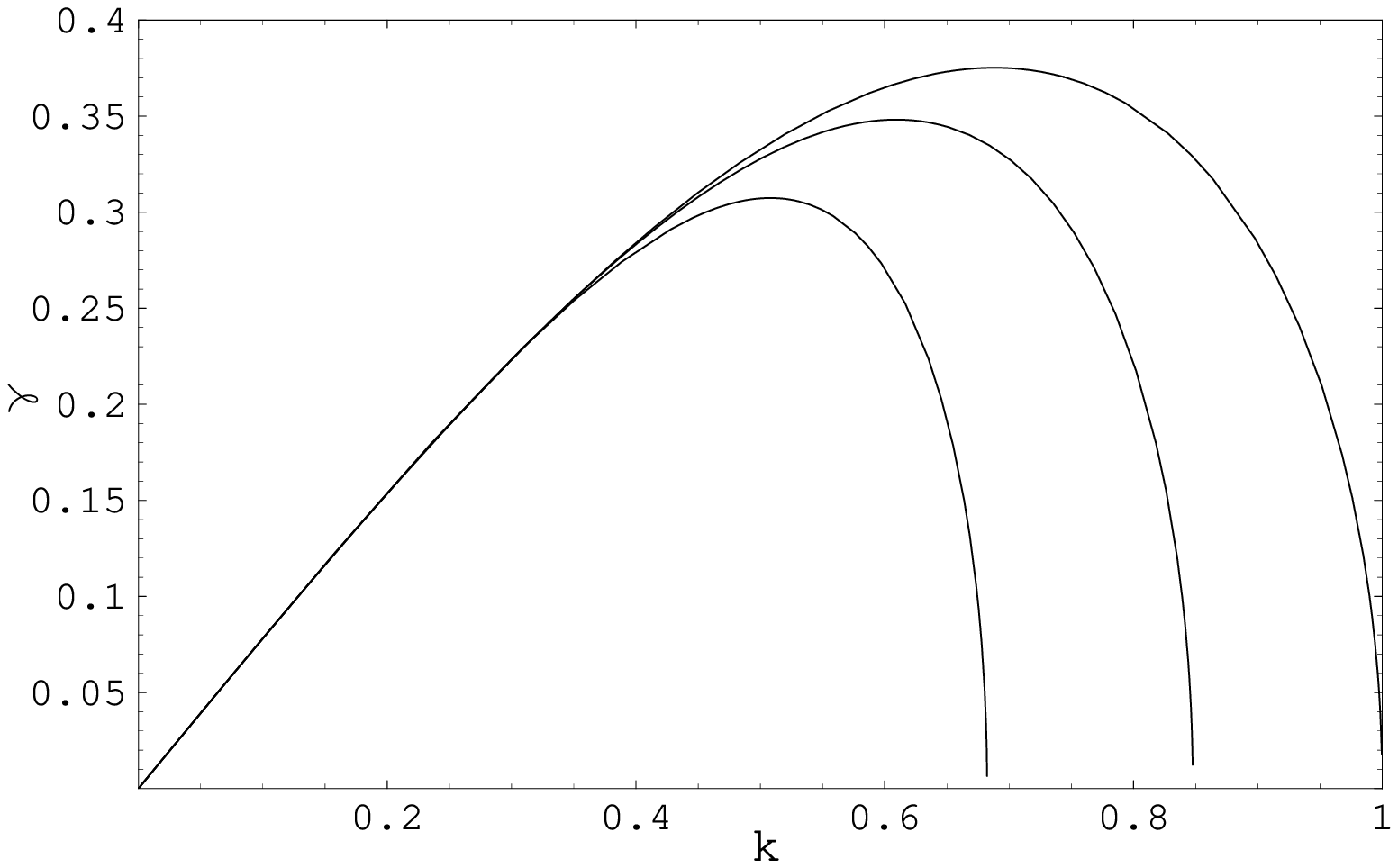}
  \caption{}
  \label{fig:1}
\end{figure}

  \noindent {Fig. 2: The classical limit of the dispersion relation 
    (\ref{eq:newdisprel}) for different values of the background
    distribution width $p_W$. We see that the upper curve in panel (a) closely 
    resembles the uppermost curve in Fig.\ \ref{fig:1}, but is slightly damped. 
    The effects of the spectral width can clearly be seen
    through the damping of the mode present in the coherent case, as well as the
    presence of a completely new mode.  
    In the different panels, we have the following widths: 
    (a) $p_W = 0.025$, 
    (b) $p_W = 0.05$, 
    (c) $p_W = 0.0667$,  
    (d) $p_W = 0.0909$, 
    and 
    (e) $p_W = 0.267$. 
    The intensity in all the panels is $I_0 = 0.50$.} \\[3mm]
 
  \noindent {Fig. 3: The classical limit of the dispersion relation 
    (\ref{eq:newdisprel}) for a select set of values of the background
    distribution width $p_W$, taken to larger wavenumbers. 
    In the different panels, we have the following widths:
    (a) $p_W = 0.0714$, 
    (b) $p_W = 0.111$,  
    and 
    (c) $p_W = 0.333$. 
    The intensity in all the panels is $I_0 = 0.50$.} \\[3mm]
    
  \noindent {Fig. 4: The inverse of the classical growth rate, i.e.\ $H = 0$, 
  plotted as a function of 
  the wavelength $\lambda = 2\pi/k$, giving the asymptotic behavior 
  of $\gamma$ for short wavelengths. Starting from the top of the panel, 
  we have used the values
  $p_W = 0.025$, 
  $p_W = 0.05$, 
  and 
  $p_W = 0.091$ 
  on the respective curve. We note the generic linear behavior for 
  short wavelengths.} \\[3mm]

  \noindent {Fig. 5 (Color online): 
  The combined effects of a quantum correction and partial coherence
    on the growth rate (black curve), obtained from Eq.\  
    (\ref{eq:newdisprel}) using $H = 0.25$, as compared to the classical case for the same
    spectral width (red curve). In the different panels, we have the following widths:
    (a) $p_W = 0.05$,  
    (b) $p_W = 0.111$, 
    and 
    (c) $p_W = 0.2$.
    We note the mode due to the quantum
    corrections combined with the partial coherence gives rise to a larger
    growth rate for short wavelengths, while the long wavelength modes are
    damped by the quantum corrections (see Fig.\ \ref{fig:1}).}

\newpage

\ \\
\newpage

\begin{figure}[ht]
  \subfigure[]{\includegraphics[width=3in]{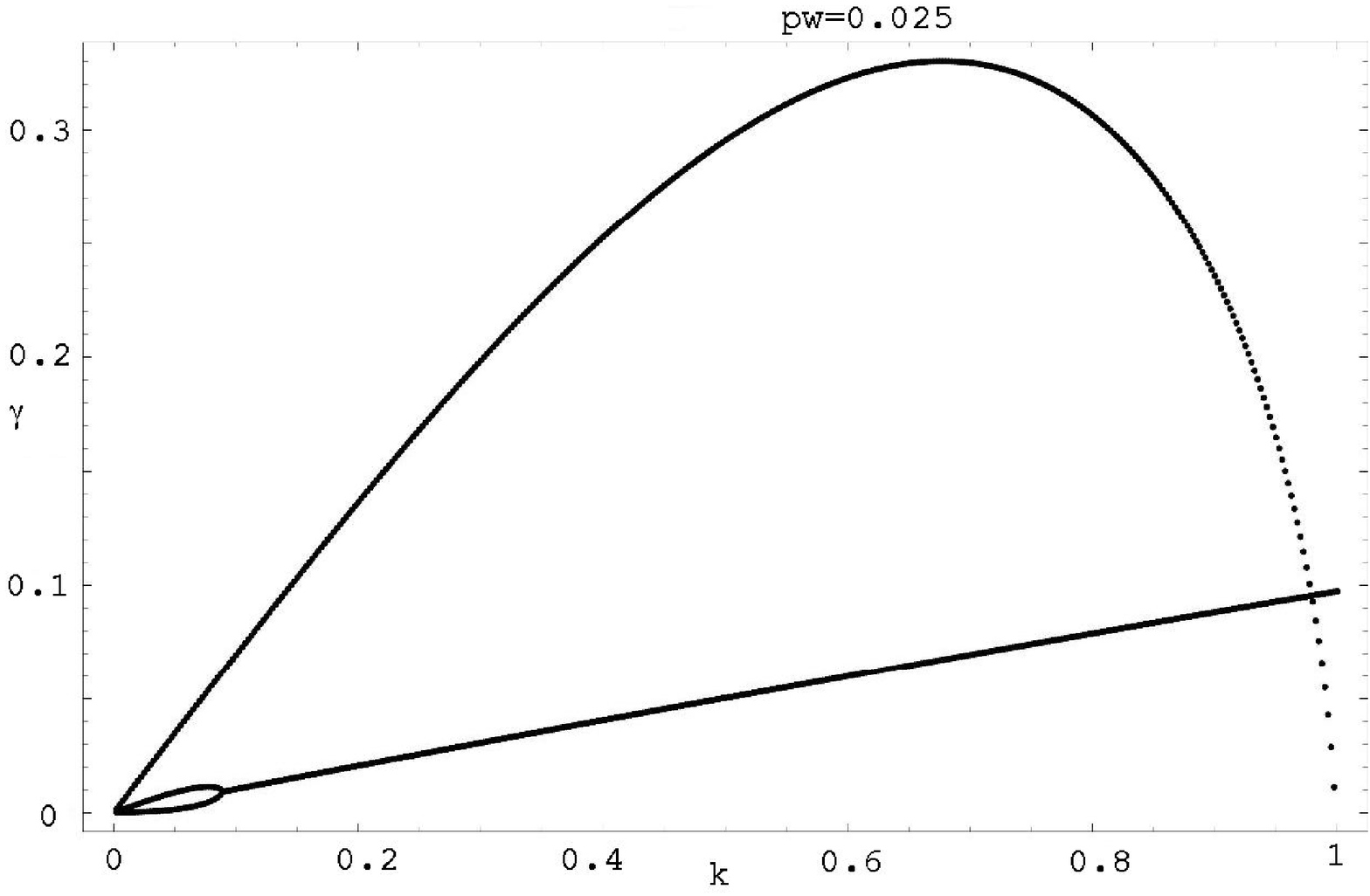}}
\end{figure}
\begin{figure}
  \subfigure[]{\includegraphics[width=3in]{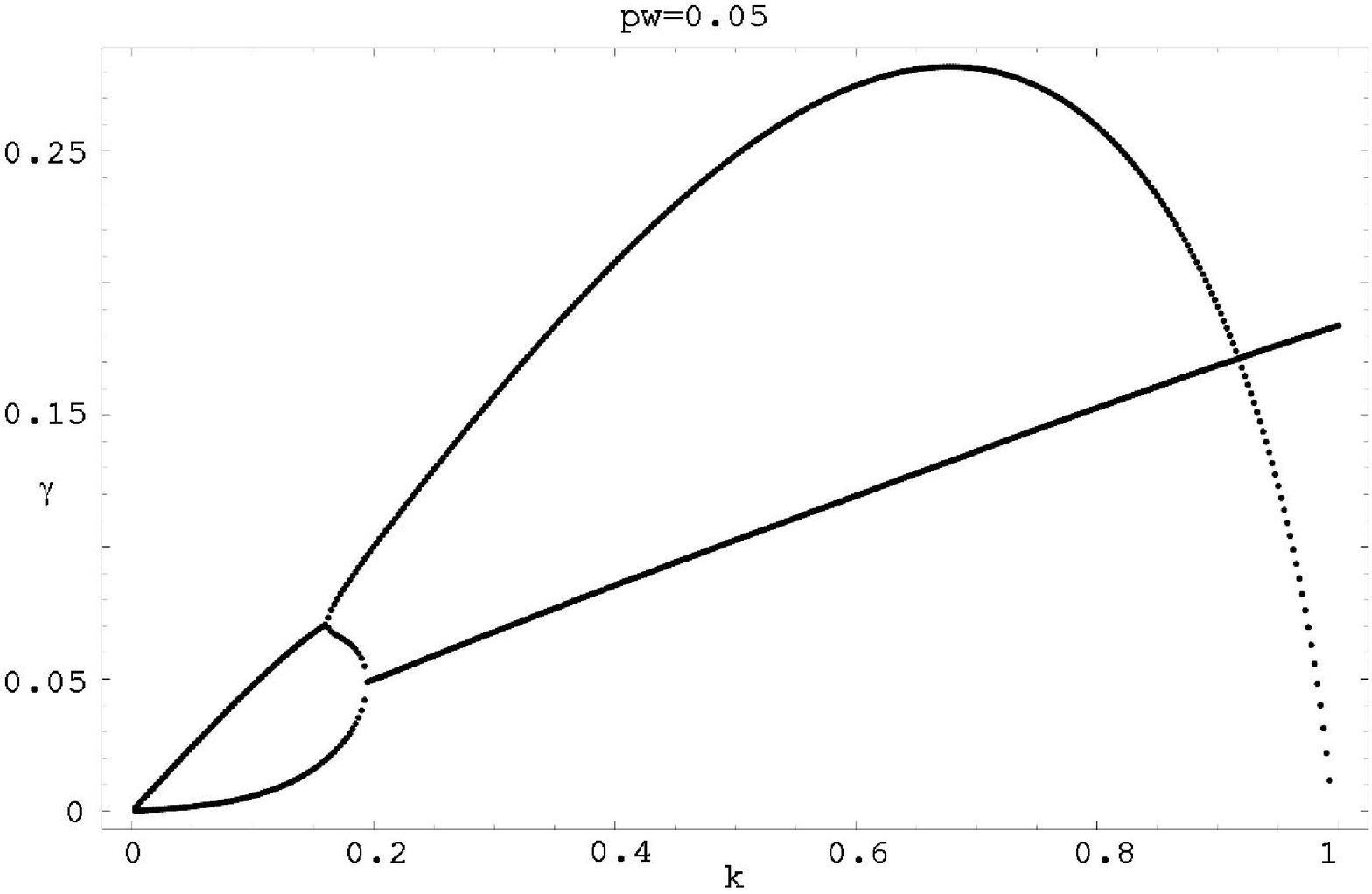}}
\end{figure}
\begin{figure}
 \subfigure[]{\includegraphics[width=3in]{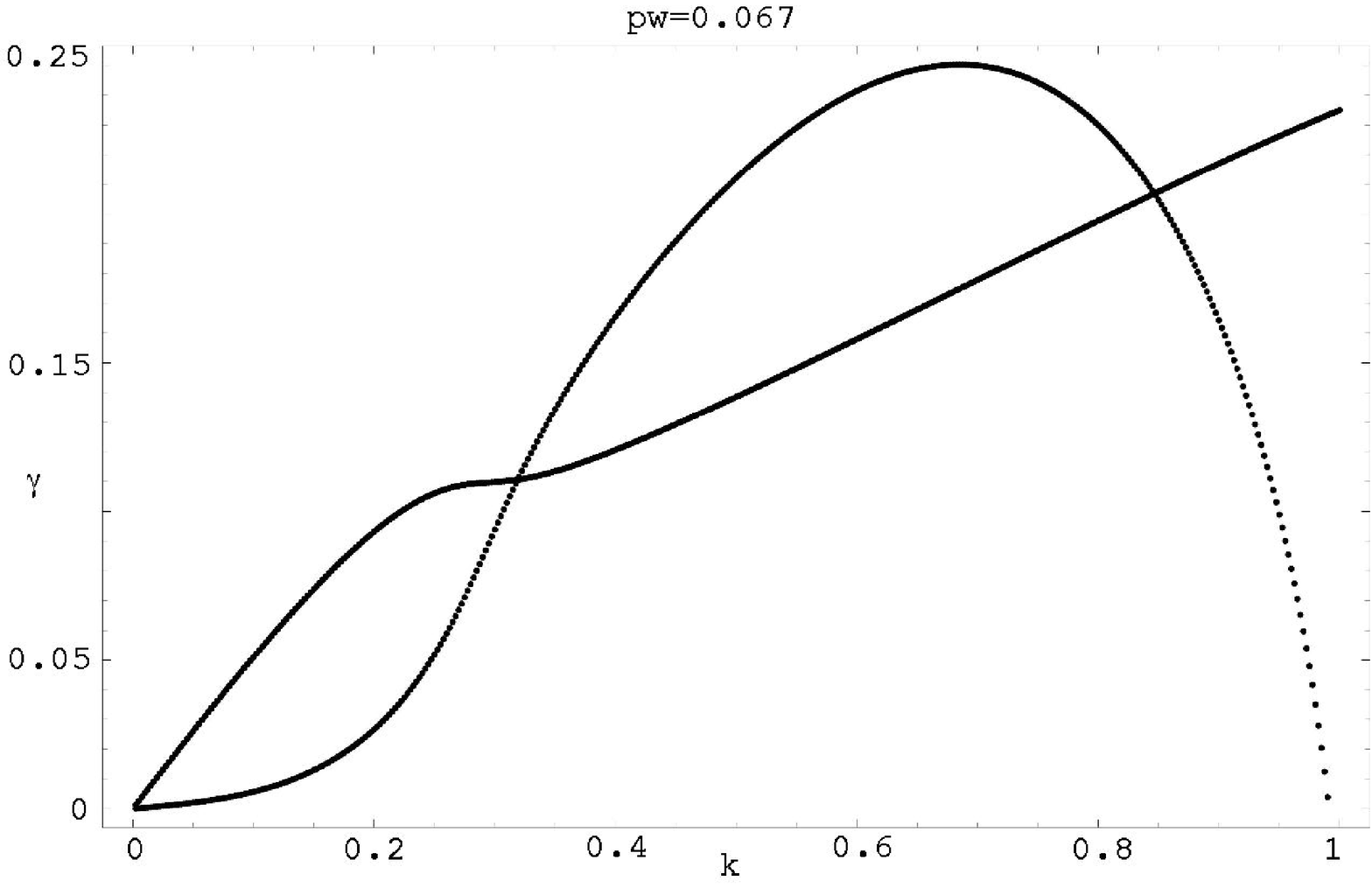}}
\end{figure}
\begin{figure}
 \subfigure[]{\includegraphics[width=3in]{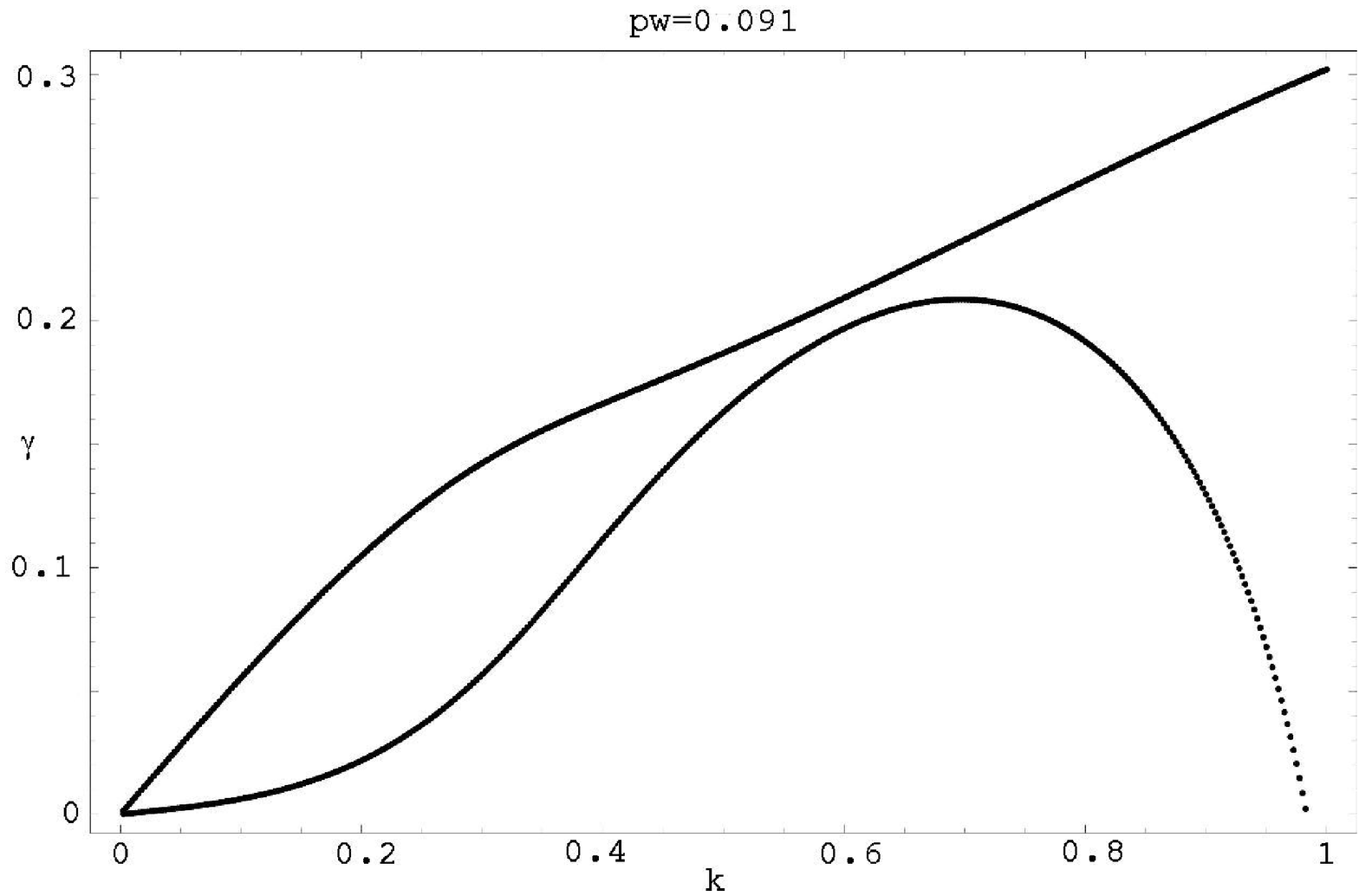}}
\end{figure}
\begin{figure}
 \subfigure[]{\includegraphics[width=3in]{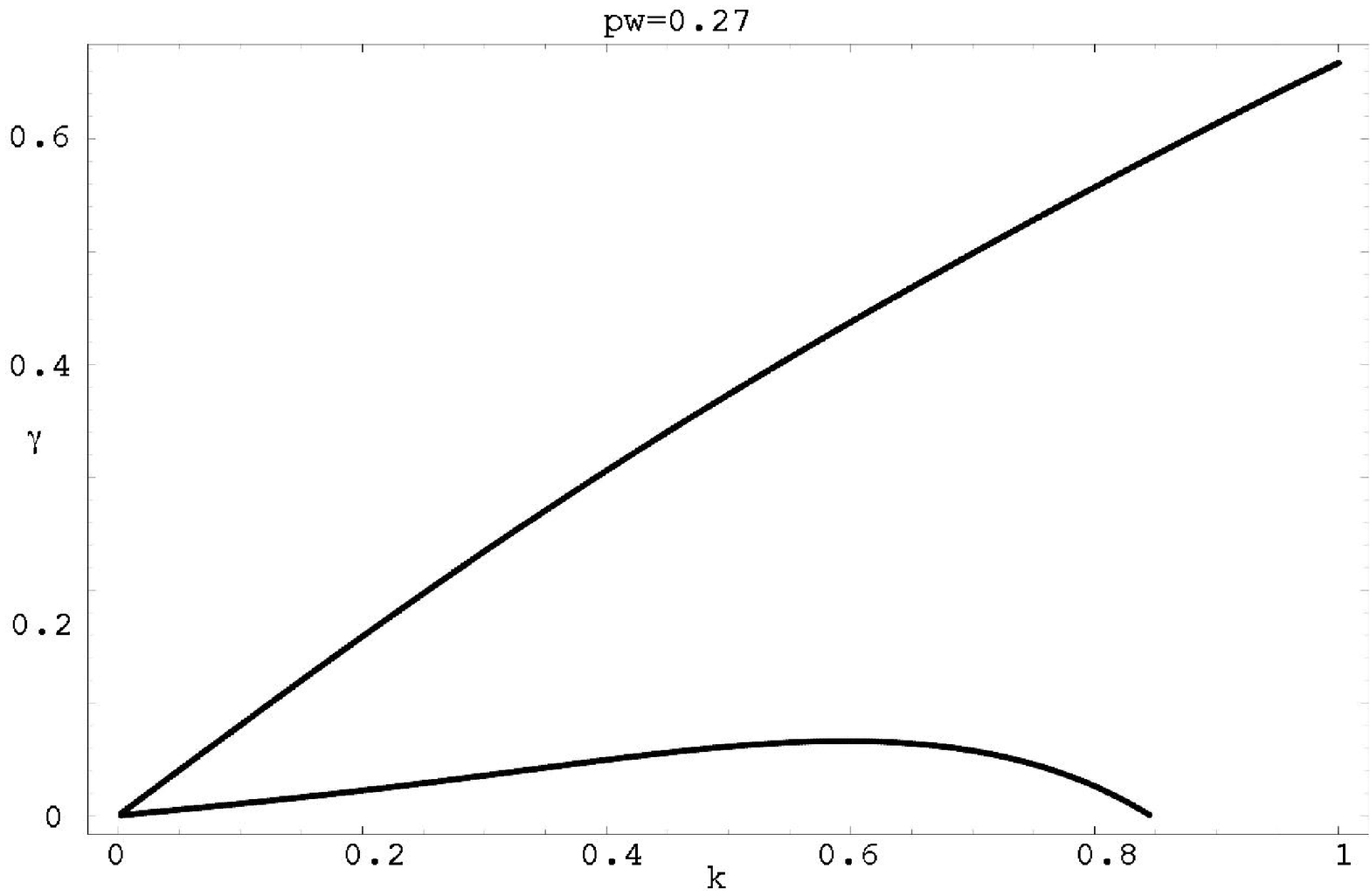}}
 \caption{}
  \label{fig:classical}
\end{figure}

\newpage

\begin{figure}[ht]
  \subfigure[]{\includegraphics[width=3in]{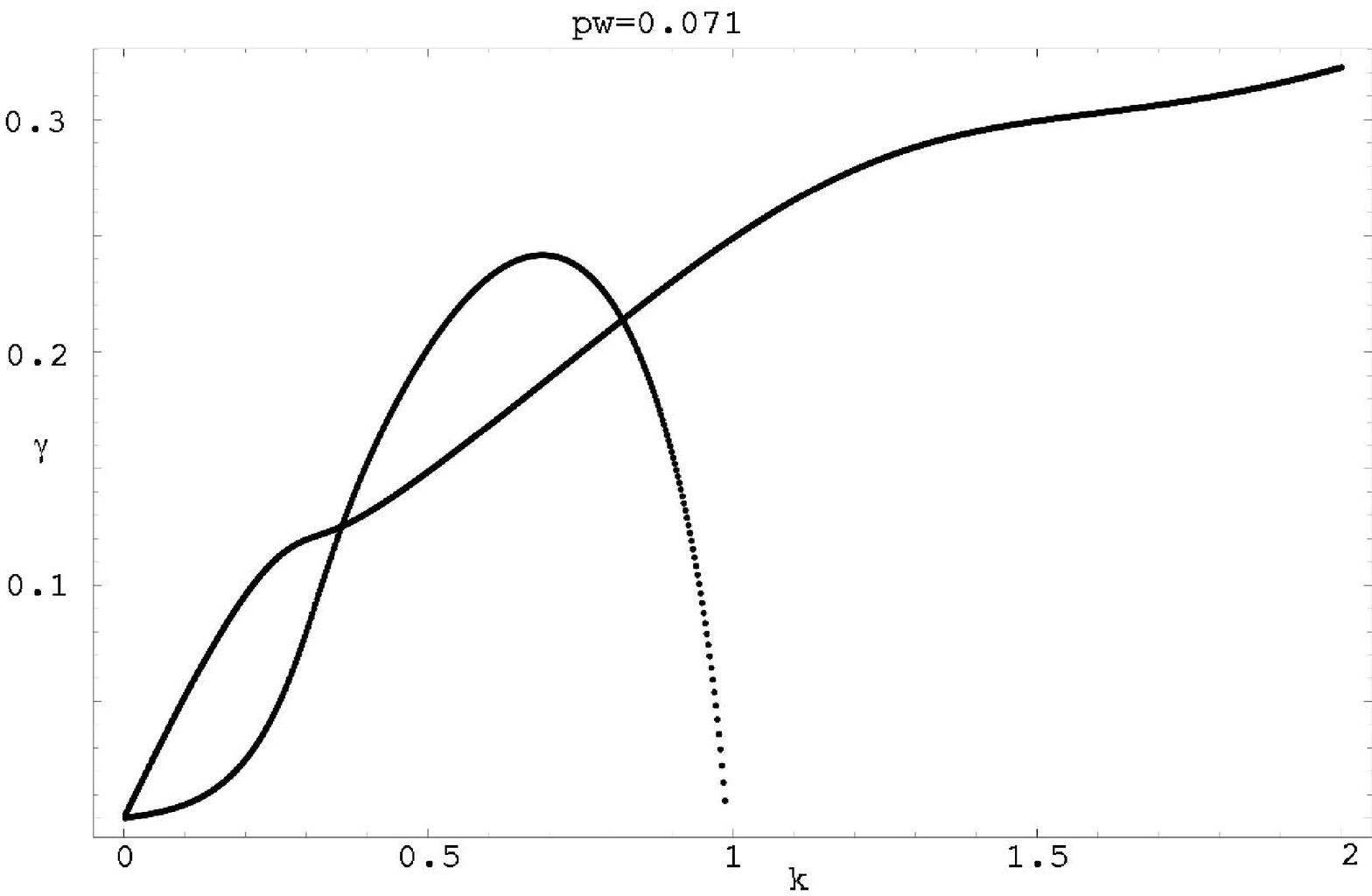}}
\end{figure}
\begin{figure}
  \subfigure[]{\includegraphics[width=3in]{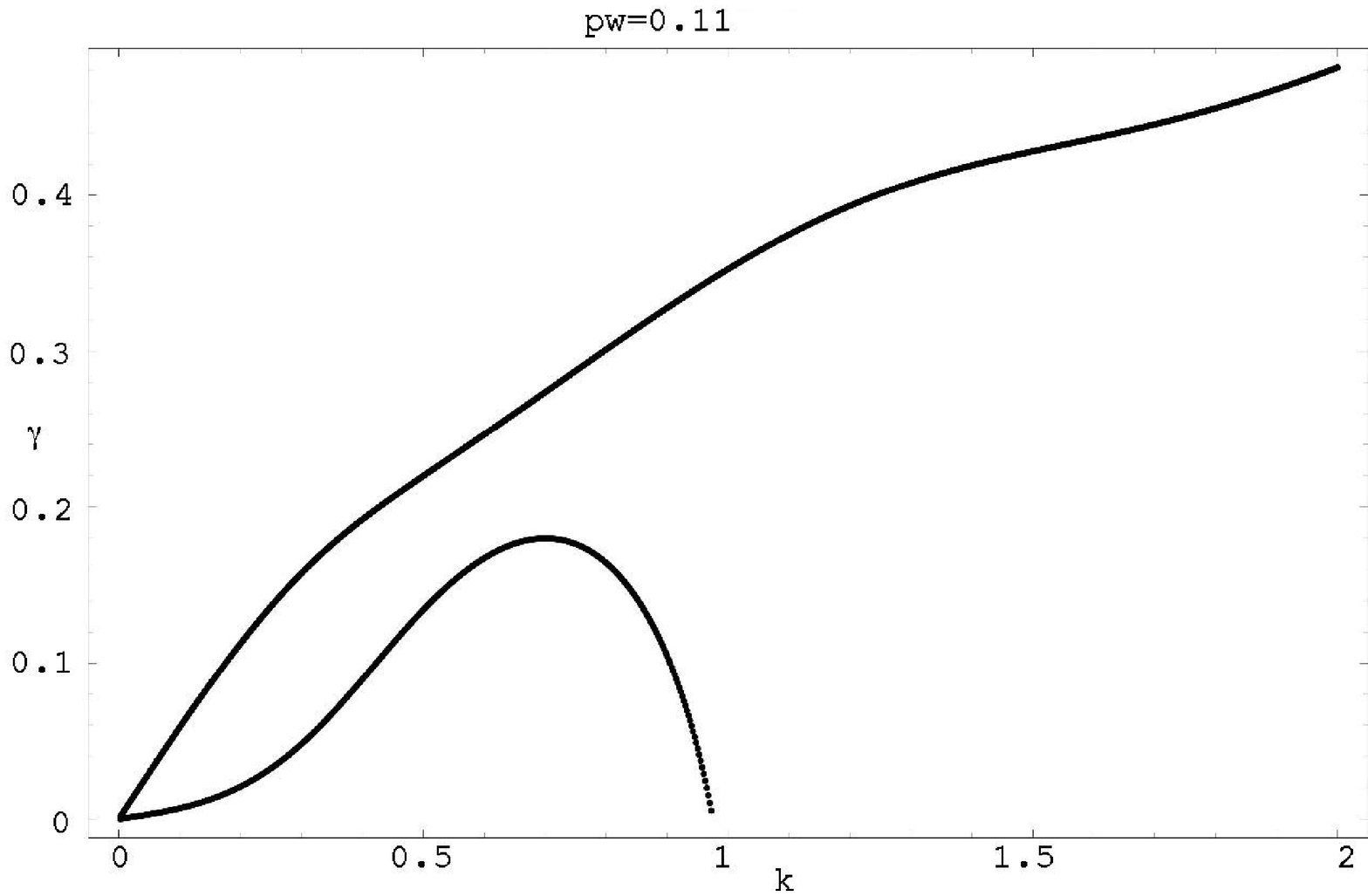}}
\end{figure}
\begin{figure}
  \subfigure[]{\includegraphics[width=3in]{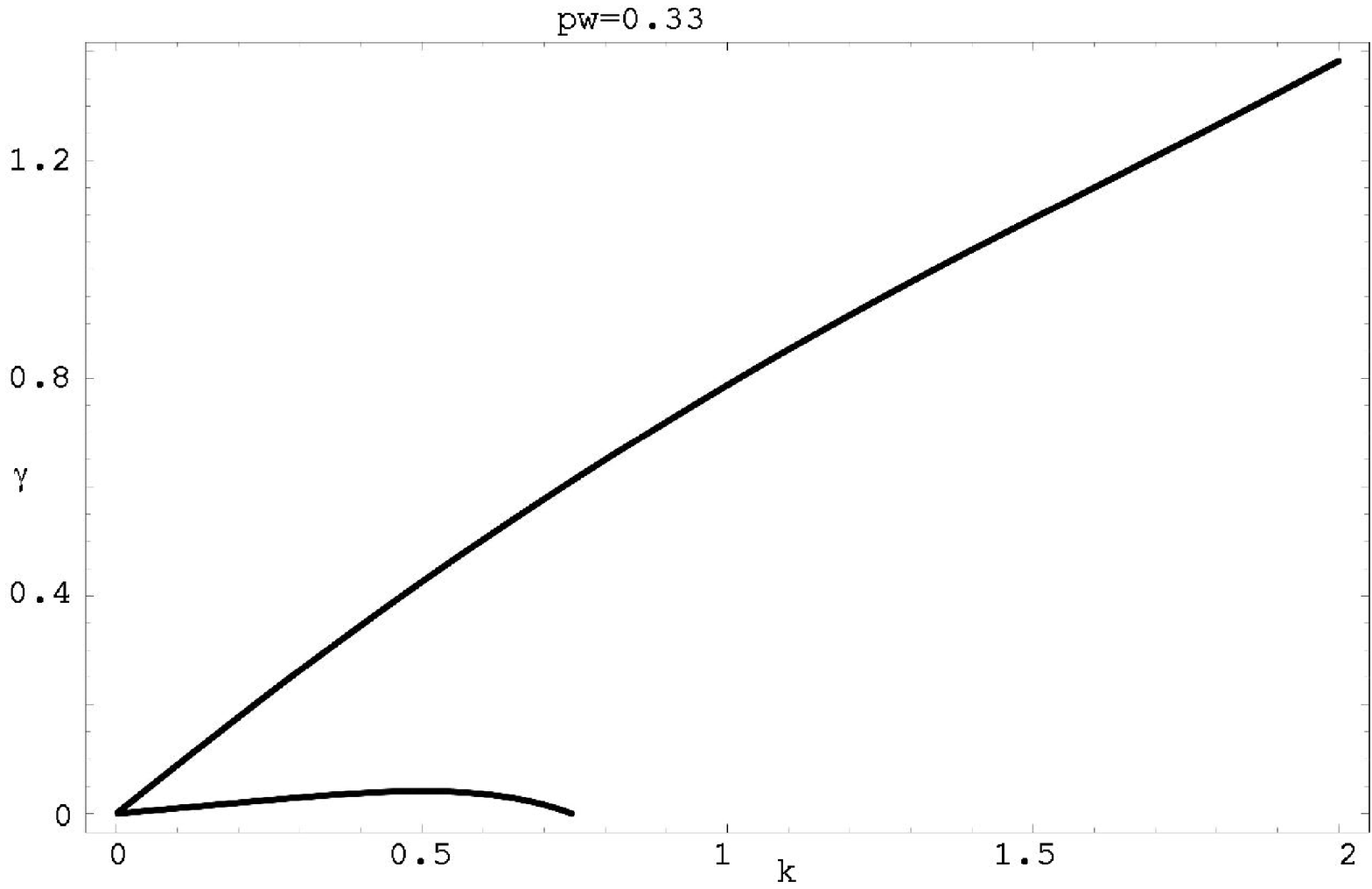}}
  \caption{}
  \label{fig:classical-large}
\end{figure}

\newpage

\begin{figure}[ht]
  \includegraphics[width=3in]{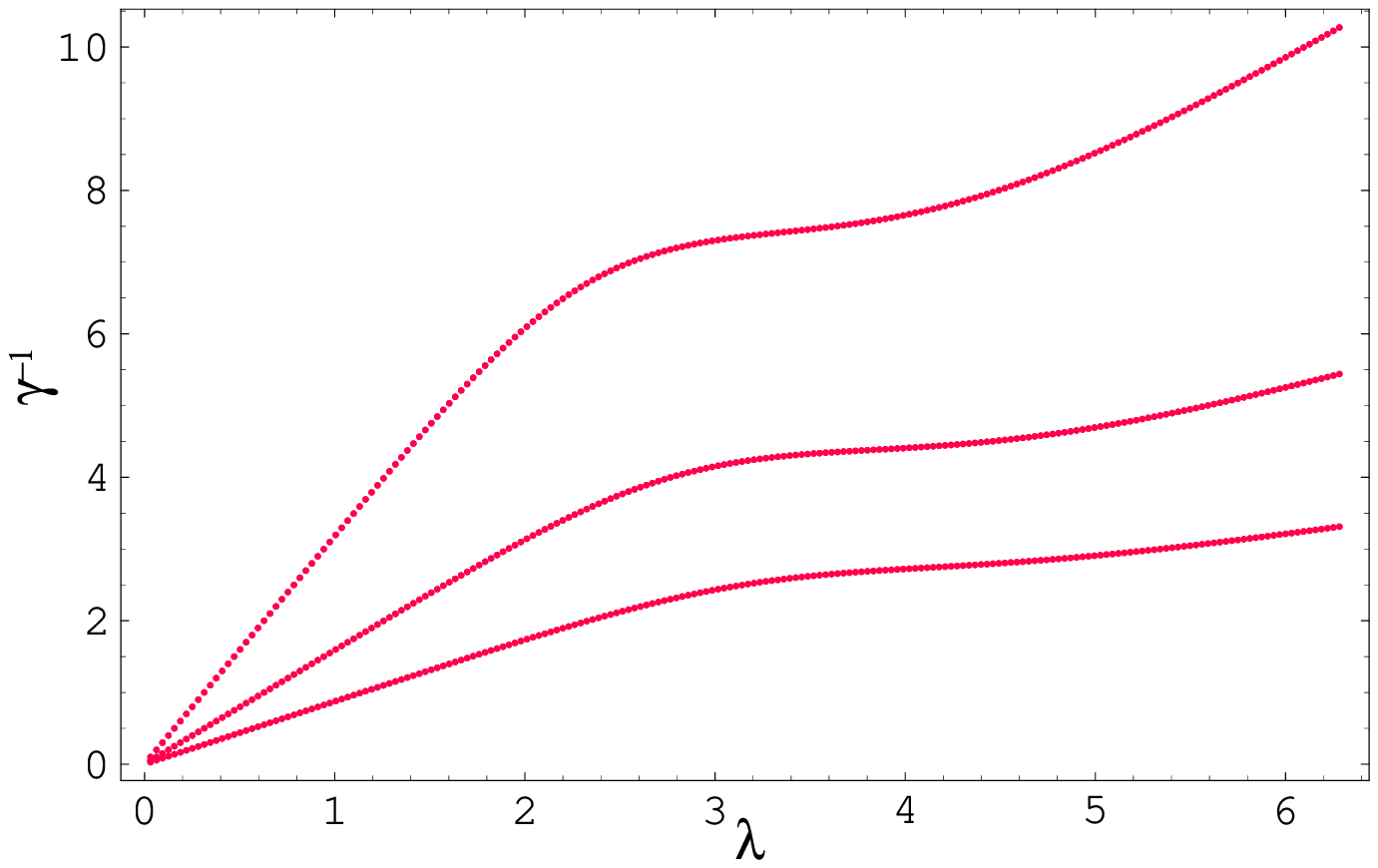}
  \caption{}
\label{fig:wavelength}
\end{figure}

\newpage

\begin{figure}
  \subfigure[]{\includegraphics[width=3in]{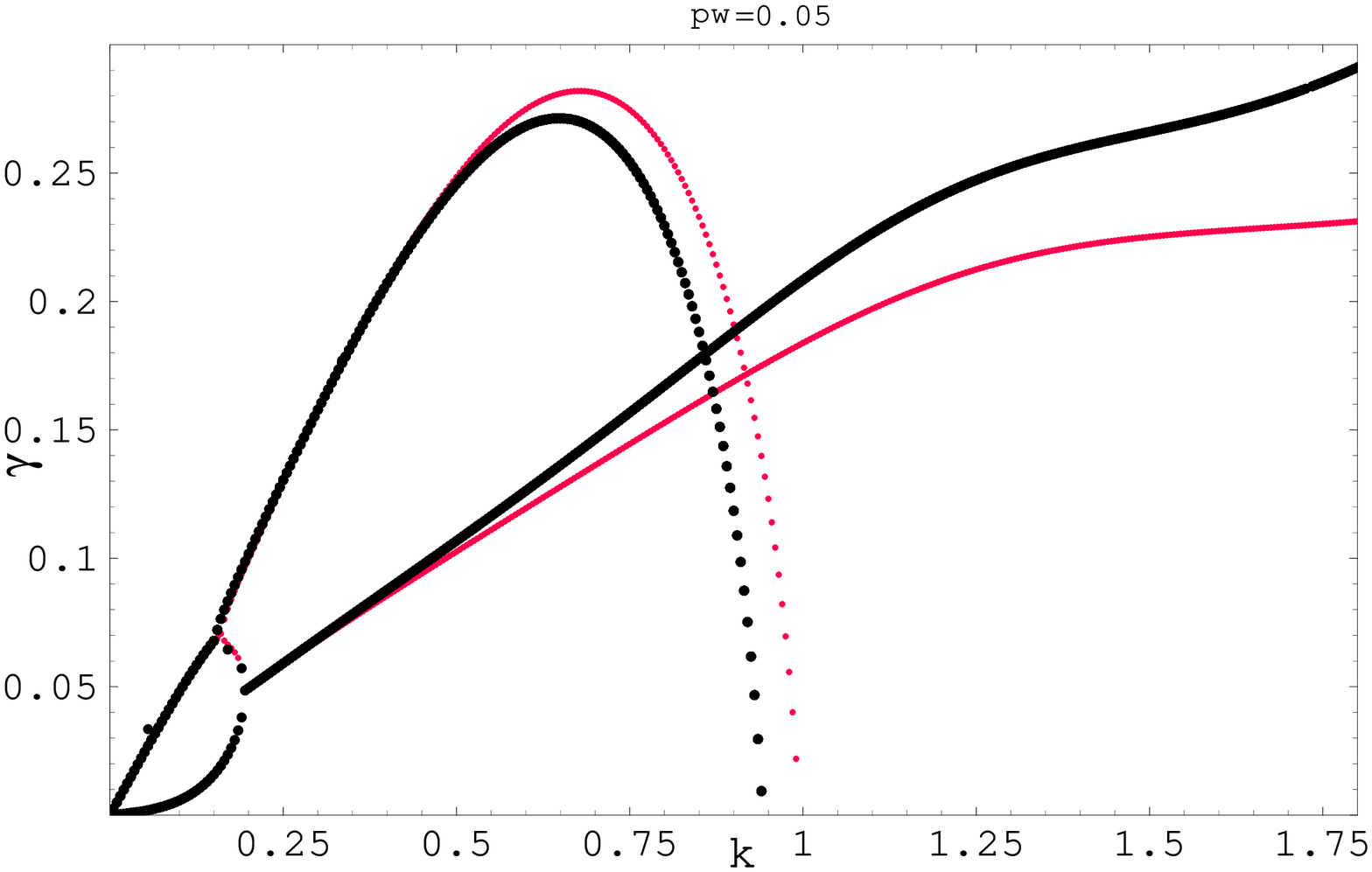}}
\end{figure}
\begin{figure}       
  \subfigure[]{\includegraphics[width=3in]{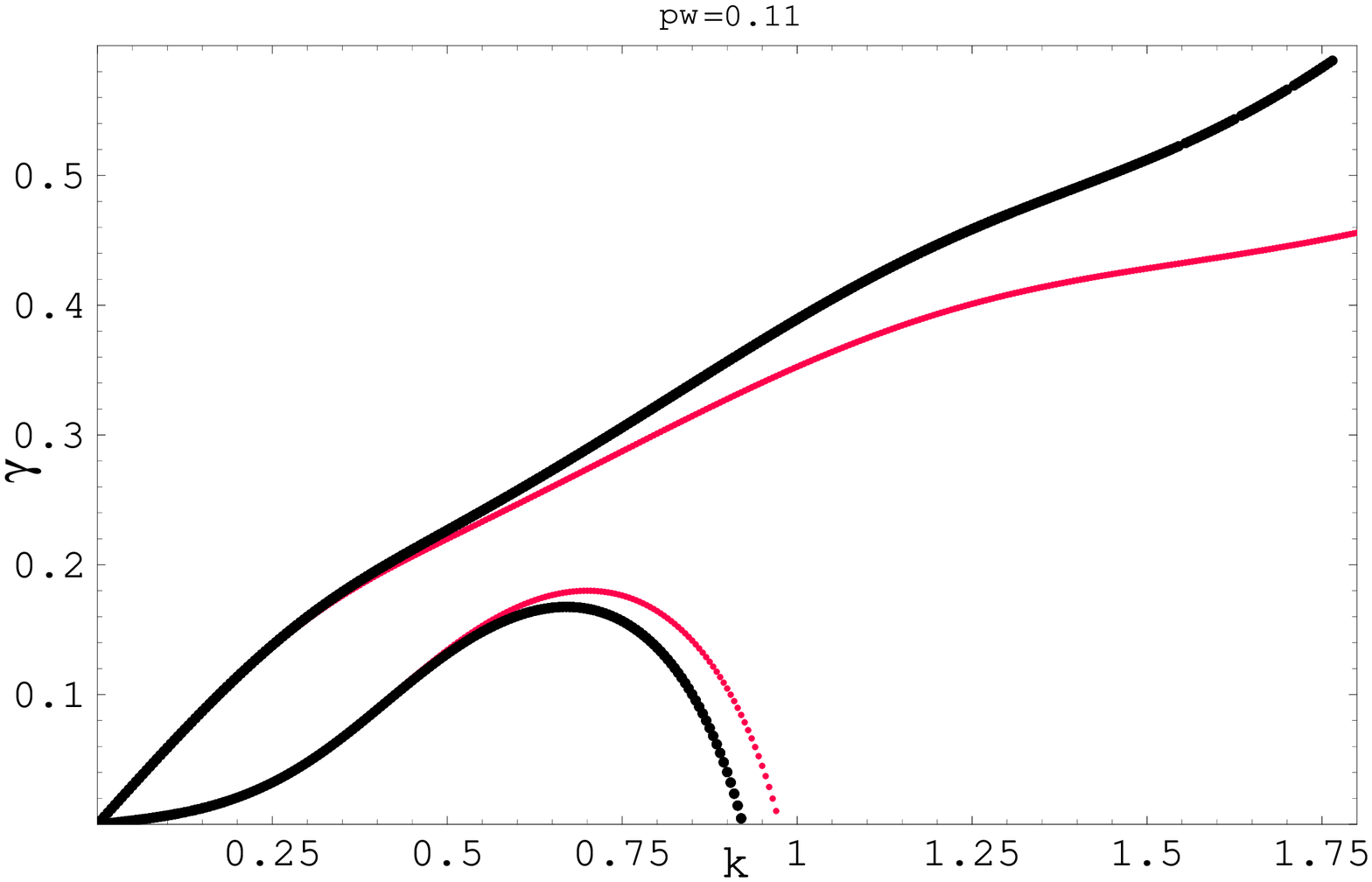}}
\end{figure}
\begin{figure}       
  \subfigure[]{\includegraphics[width=3in]{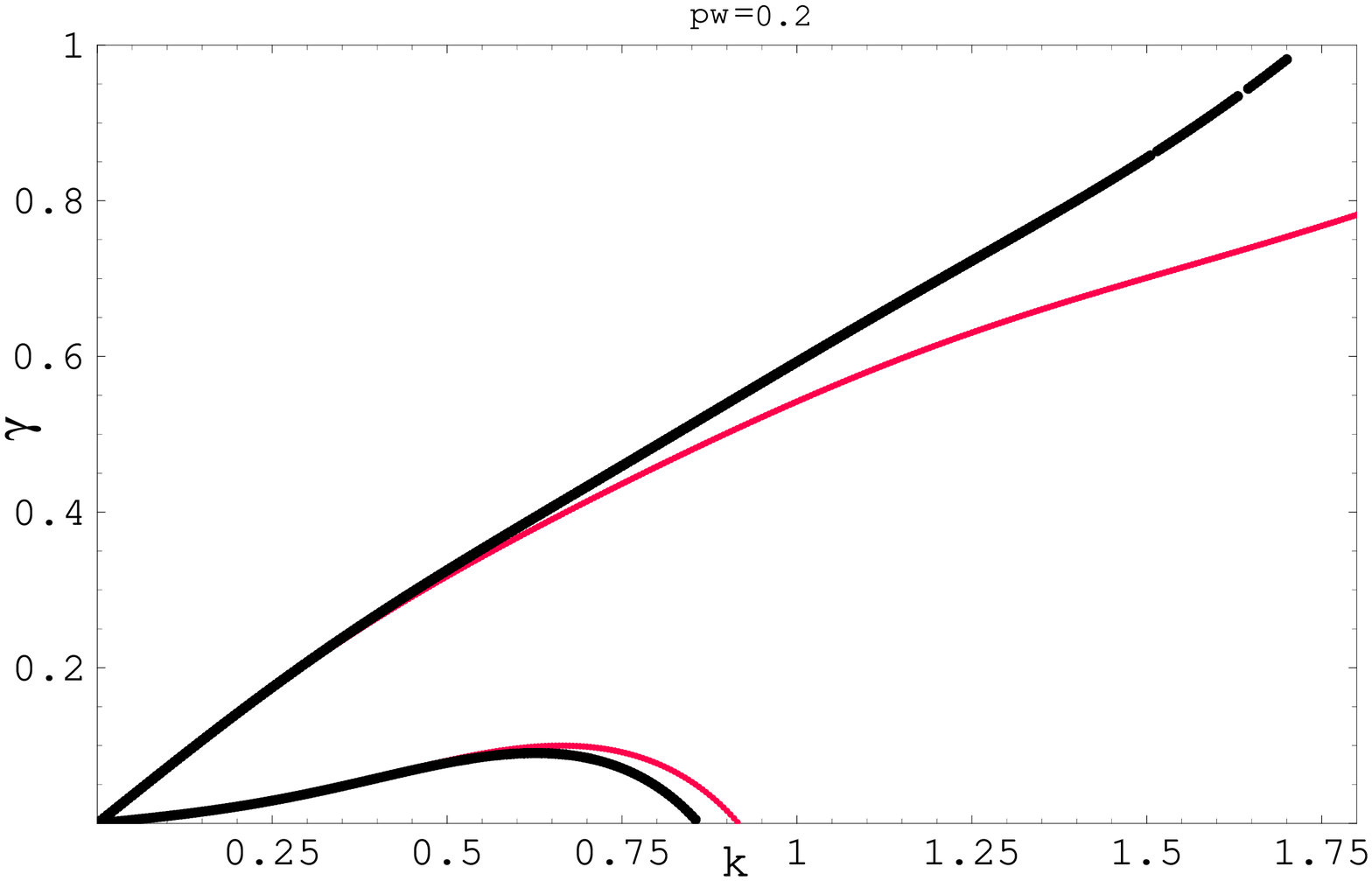}}       
  \caption{}
  \label{fig:quantum}
\end{figure}

\end{document}